\def\papertitle{TorchFX: A Modern Approach to Audio DSP with PyTorch and GPU Acceleration}
\def\paperauthorA{Matteo Spanio}
\def\paperauthorB{Antonio Rodà}

\documentclass[twoside,a4paper]{article}
\usepackage{etoolbox}
\usepackage{dafx_25}
\usepackage{amsmath,amssymb,amsfonts,amsthm}
\usepackage{euscript}
\usepackage[T1]{fontenc}
\usepackage[utf8]{inputenc}
\usepackage{ifpdf}
\usepackage[english]{babel}
\usepackage{caption}
\usepackage{subcaption} 
\usepackage{color}
\usepackage{booktabs}

\input glyphtounicode
\ninept

\newcounter{numauth}
\newcounter{listcnt}
\newcommand\authcnt[1]{\ifdefined#1 \stepcounter{numauth} \fi}

\newcommand\addauth[1]{
\ifdefined#1 
\stepcounter{listcnt}
\ifnum \value{listcnt}<\value{numauth}
\appto\authorslist{, #1}
\else
\appto\authorslist{~and~#1}
\fi
\fi}
\authcnt{\paperauthorB}
\authcnt{\paperauthorC}
\authcnt{\paperauthorD}
\authcnt{\paperauthorE}
\authcnt{\paperauthorF}
\authcnt{\paperauthorG}
\authcnt{\paperauthorH}
\authcnt{\paperauthorI}
\authcnt{\paperauthorJ}
\def\authorslist{\paperauthorA}
\addauth{\paperauthorB}
\addauth{\paperauthorC}
\addauth{\paperauthorD}
\addauth{\paperauthorE}
\addauth{\paperauthorF}
\addauth{\paperauthorG}
\addauth{\paperauthorH}
\addauth{\paperauthorI}
\addauth{\paperauthorJ}

\usepackage{times}

\newif\ifpdf
\ifx\pdfoutput\relax
\else
   \ifcase\pdfoutput
      \pdffalse
   \else
      \pdftrue
   \fi
\fi

\ifpdf 
  \usepackage[pdftex,
    pdftitle={\papertitle},
    pdfauthor={\authorslist},
    pdfsubject={Proceedings of the 28th International Conference on Digital Audio Effects (DAFx25)},
    colorlinks=false, 
    bookmarksnumbered, 
    pdfstartview=XYZ 
  ]{hyperref}
  \usepackage[pdftex]{graphicx}
\else 
  \usepackage[dvips]{epsfig,graphicx}
  \usepackage[dvips,
    pdftitle={\papertitle},
    pdfauthor={\authorslist},
    pdfsubject={Proceedings of the 28th International Conference on Digital Audio Effects (DAFx25)},
    colorlinks=false, 
    bookmarksnumbered, 
    pdfstartview=XYZ 
  ]{hyperref}
\fi
\usepackage[hypcap=true]{caption}
\title{\papertitle}

\affiliation{
\paperauthorA\ and \paperauthorB\,\thanks{\vspace{-3mm}}}
{\href{https://csc.dei.unipd.it/}{Centro di Sonologia Computazionale (CSC)} \\ Dept. of Information Engineering \\ University of Padova \\ Padova, IT\\
{\tt \href{mailto:spanio@dei.unipd.it}{\{spanio,roda\}@dei.unipd.it}}
}

\begin{document}
\ifpdf 
  \DeclareGraphicsExtensions{.png,.jpg,.pdf}
\else  
  \DeclareGraphicsExtensions{.eps}
\fi


\maketitle

\begin{abstract}
The burgeoning complexity and real-time processing demands of audio signals necessitate optimized algorithms that harness the computational prowess of Graphics Processing Units (GPUs). Existing Digital Signal Processing (DSP) libraries often fall short in delivering the requisite efficiency and flexibility, particularly in integrating Artificial Intelligence (AI) models. In response, we introduce TorchFX: a GPU-accelerated Python library for DSP, specifically engineered to facilitate sophisticated audio signal processing. Built atop the PyTorch framework, TorchFX offers an Object-Oriented interface that emulates the usability of torchaudio, enhancing functionality with a novel pipe operator for intuitive filter chaining. This library provides a comprehensive suite of Finite Impulse Response (FIR) and Infinite Impulse Response (IIR) filters, with a focus on multichannel audio files, thus facilitating the integration of DSP and AI-based approaches. Our benchmarking results demonstrate significant efficiency gains over traditional libraries like SciPy, particularly in multichannel contexts. Despite current limitations in GPU compatibility, ongoing developments promise broader support and real-time processing capabilities. TorchFX aims to become a useful tool for the community, contributing to innovation and progress in DSP with GPU acceleration. TorchFX is publicly available on GitHub at \url{https://github.com/matteospanio/torchfx}.
\end{abstract}

\section{Introduction}
\label{sec:intro}

As the applications of DSP to fields such as telecommunications, multimedia, and artificial intelligence (AI) become more complex and the demand for real-time processing intensifies, the necessity for optimized algorithms that harness the computational prowess of Graphics Processing Units (GPUs) becomes paramount.

Despite the strides made in GPU technology, existing DSP libraries frequently fall short in delivering the requisite efficiency and flexibility demanded by contemporary applications. These libraries, while operational, often possess design constraints that impede the full exploitation of GPU capabilities \cite{2020SciPy-NMeth, harris2020array}. Moreover, the integration of AI models into DSP workflows necessitates a seamless and intuitive interface capable of accommodating the complexities inherent in both domains. The absence of such a library creates a void within the ecosystem, compelling researchers and practitioners to navigate cumbersome interfaces or develop bespoke solutions that may not fully leverage GPU acceleration when developing complex AI systems.

In response to this critical need, we present TorchFX, an innovative Python library constructed atop the PyTorch framework, specifically engineered to facilitate GPU-accelerated DSP for audio signals. This library is designed to provide an Object-Oriented interface that emulates the usability of torchaudio, ensuring compatibility while enhancing functionality. By prioritizing user experience, TorchFX incorporates operator overloading, enabling the intuitive chaining of filters through the use of the bitwise OR (|) operator. This design choice not only simplifies the construction of complex audio processing pipelines but also aligns with the burgeoning trend of modularity in software design, as exemplified by frameworks like Langchain \cite{Chase_LangChain_2022} in natural language processing (NLP).

TorchFX offers a comprehensive suite of filters, including both Finite Impulse Response (FIR) and Infinite Impulse Response (IIR) filters, to address the diverse requirements of audio manipulation, with a particular focus on multichannel audio files. This capability is indispensable in today’s multimedia environment, where audio content frequently spans multiple channels and necessitates precise processing to achieve the desired output. By bridging the gap between DSP and AI, TorchFX aims to empower researchers and developers by facilitating the implementation of advanced audio processing techniques.

The structure of this paper is organized to provide an understanding of TorchFX and its contributions to the DSP landscape. Section \ref{sec:background} presents a comprehensive survey of the background, highlighting the evolution of GPU-accelerated libraries and existing DSP solutions. In Section \ref{sec:design}, we delve into the design principles and core features of TorchFX, elucidating how it addresses the limitations of current libraries. Section \ref{sec:benchmark} provides a thorough evaluation of the library, demonstrating the efficiency of TorchFX and showcasing its capabilities through real-world examples. Finally, Section \ref{sec:conclusion} concludes the paper by summarizing our findings and outlining potential directions for future research and development.

\section{Background and related works}
\label{sec:background}

In the realm of Digital Signal Processing (DSP) for audio, the introduction of a new tool into the already diverse landscape of Python libraries presents both challenges and opportunities. Existing solutions, while varied, often fail to comprehensively address all potential use cases, particularly in terms of functionality and tool design. To elucidate the current state of the art and identify gaps, we conducted a literature review focusing on Python libraries for audio DSP. The primary aim of this review was to assess the landscape of available tools and their capabilities, particularly concerning GPU acceleration, which is increasingly important for handling complex audio processing tasks.

The research questions guiding our investigation were meticulously crafted to provide a comprehensive understanding of the current offerings:

\begin{itemize}
\item How many GPU-accelerated libraries have been released for Python?
\item What functionalities do these libraries offer?
\item How is their interface designed?
\end{itemize}

\begin{figure}[ht]
    \centering
    \includegraphics[width=\linewidth]{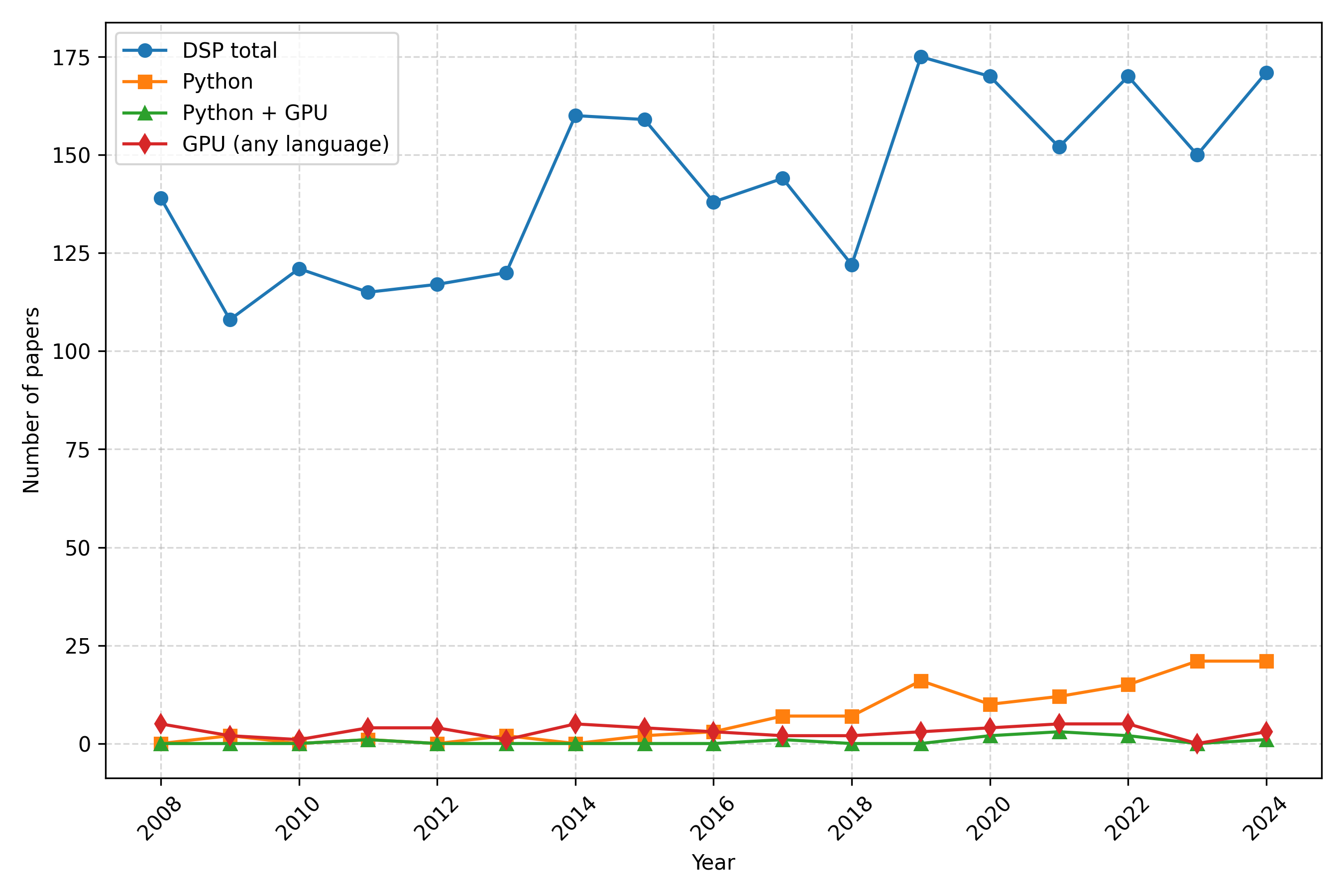}
    \caption{Scientific Literature on DSP Software (2008---2024)}
    \label{fig:lit-rev}
\end{figure}

To address these questions, we utilized the Scopus search engine to perform a systematic literature review. The inclusion criteria were defined using the search query: TITLE-ABS-KEY ( ( ( signal AND processing ) OR dsp ) AND ( library OR package OR software ) AND ( audio OR sound OR music ) ) AND PUBYEAR > 2007 AND PUBYEAR < 2025. The starting point of 2008 was chosen to coincide with the first stable release of Python 3.0, marking a significant evolution in Python’s capabilities, while 2025 was excluded as it is yet to conclude. The inclusion of terms such as ``library,'' ``package,'' and ``software'' was strategic, ensuring a focus on software tool implementations and excluding generic DSP processes without a designated interface for further development by other programmers.

Our search yielded 2431 papers on digital signal processing libraries for audio published between 2008 and 2024. Of these, 119 were available in Python, with only 10 primarily focusing on GPU acceleration \cite{lee_grafx_2024, Yang2022Torchaudio, Diaz-Guerra20215653, cheuk_nnaudio_2020, Ahmed2021, ramaiah_dynamic_2020, bucurica_improving_2017, fontaine_brian_2011}. This indicates a relatively limited focus on leveraging GPU capabilities within the Python DSP community.

Subsequently, we refined our selection by excluding articles not directly related to audio DSP or the specific development of libraries, whether GPU or CPU-based. The final selection comprised papers associated with the most widely used libraries in the audio domain, as summarized in Table \ref{tab:audio-libraries-merged}. The table provides a detailed overview of the libraries, their associated papers and authors, and key characteristics such as GPU acceleration, the level of the API (high or low), the availability of explicit filter implementations, and the capability for signal analysis.

\begin{table*}[ht]
\centering
\caption{\itshape Overview and Features of Major Audio Libraries.}
\begin{tabular}{|c|c|c|c|c|c|c|}
\hline
Authors & Paper & Library & Filters & API & Signal Analysis & GPU \\
\hline
Cheuk et al. & \cite{cheuk_nnaudio_2020} & nnAudio & no & low-level & no & yes \\
Yang et al. & \cite{Yang2022Torchaudio, torch2} & torchaudio & yes & low-level & yes & yes \\
McFee et al. & \cite{mcfee_librosa_2015} & librosa & no & low-level & yes & no \\
Virtanen et al. & \cite{2020SciPy-NMeth} & scipy & yes & high-level & yes & no \\
\hline
\end{tabular}
\label{tab:audio-libraries-merged}
\end{table*}

The analysis of this survey reveals a notable lack of interest within the scientific community in releasing efficient libraries for signal processing, with a tendency to develop ad-hoc solutions that are not adequately shared or valorized. Considering the substantial body of research related to DSP, it would be prudent to systematically collect and share such software with the broader community. Furthermore, none of the solutions identified in the survey provide an object-oriented interface, instead relying on interfaces derived from MATLAB functions \cite{MATLAB}. This highlights a significant gap in the availability of user-friendly, modular, and extensible DSP libraries that can fully exploit the capabilities of modern computing architectures, such as GPUs, while offering an intuitive development experience.

\section{Design principles and core features}
\label{sec:design}

Inspired by the design paradigm established by torchaudio \cite{Yang2022Torchaudio}, which is widely regarded as the standard for deep learning applications involving audio, our objective was to implement a distinct class for each filter type to ensure seamless compatibility with \texttt{torchaudio}.

To promote a more object-oriented methodology for audio manipulation, we encapsulated both the signal and its sampling frequency within a single class, \texttt{Wave}. Within this class, we overloaded the OR operator to function similarly to a pipe operator, akin to that used in Bash. This design choice marks a departure from the conventional representation of audio signals in existing libraries, which typically provide the sample array and sampling frequency as separate entities, with the frequency being utilized only during signal manipulation. Given that numerous discrete algorithms necessitate knowledge of the sample rate, the approach adopted by \texttt{torchfx} (similar to that proposed by the \texttt{pydub} library) ensures that these discrete parameters remain transparent and consistent throughout the program's execution. This design also mitigates potential bugs arising from omitted parameters; for example, the \texttt{librosa} \cite{mcfee_librosa_2015} library often defaults to a sampling frequency of 22050 Hz. Additionally, the creation of the class facilitates operator overloading and the inclusion of supplementary methods.

In \texttt{torchfx}, we have leveraged the strengths of \texttt{torchaudio} and extended its interface. Essentially, the \texttt{Transforms} library within \texttt{torchaudio} offers several implementations of transformations applicable to signals (e.g., \texttt{Resample} for altering the sampling rate or \texttt{Vol} for adjusting signal gain), alongside enabling transitions between the time domain and frequency domain using algorithms such as Short-Time Fourier Transform (STFT), Mel spectrogram extraction, and the Griffin-Lim algorithm.

The section pertaining to IIR and FIR filters, however, does not implement a similar interface and is instead confined to the \texttt{functionals} module of \texttt{torchaudio}, which emulates a MATLAB-style interface\footnote{An additional interface is provided by the sox bindings, but it lacks GPU acceleration.}. This module does not offer default implementations of various filter types (e.g., Butterworth, peaking, Chebyshev, shelving) but instead provides a single generic function, \texttt{lfilter}, which requires the user to supply filter coefficients. Our aim was to develop a foundational implementation of various FIR and IIR filter types through an object-oriented interface analogous to the transformations provided by \texttt{torchaudio}, thereby ensuring compatibility with modules used for constructing neural networks, based on the \texttt{nn.Module} class provided by PyTorch.

Moreover, we sought to simplify and streamline the construction of processing pipelines by overloading the bitwise OR operator. This approach is particularly advantageous as creating classes often necessitates writing boilerplate code, which is typically avoided, especially when dealing with complex filter chains or when the implementation of a neural network is not desired.

The following code snippet demonstrates the potential interfaces for implementing a simple filter chain:
\begin{verbatim}
from torch import nn
from torch.nn import Sequential
from torchfx.signal import Wave
from torchfx.filter import (
    HiShelving,
    LoShelving,
)

signal = Wave.from_file("path_to_file.mp3")

# Implementation using classes:
class FilterChain(nn.Module):
    def __init__(self, sample_rate):
        super().__init__(self)
        self.f1 = HiShelving(1000, sample_rate)
        self.f2 = LoShelving(2000, sample_rate)

    def forward(self, x):
        x = self.f1(x)
        x = self.f2(x)
        return x

fchain = FilterChain(signal.fs)
result = fchain(signal.y)

# Implementation using Sequential
fchain = Sequential([
    HiShelving(1000, sample_rate=signal.fs),
    LoShelving(2000, sample_rate=signal.fs),
])
result = fchain(signal.y)

# Implementation using pipe operator
result = signal \
  | HiShelving(1000) \
  | LoShelving(2000)
\end{verbatim}
In the final example, the sample rate of the discrete filters is omitted, as it is lazily evaluated during filter application, thanks to the pipe operator. Alternatively, one can define the filter chain within \texttt{Sequential} without specifying the sampling frequency and subsequently apply the filters to the signal using the pipe operator. The operator overloading also manages the sampling frequency, rendering the process transparent:

\begin{verbatim}
# Implementation using Sequential
fchain = Sequential([
    HiShelving(1000), 
    LoShelving(2000),
])
result = signal | fchain
\end{verbatim}

\section{Performance Evaluation}
\label{sec:benchmark}

To evaluate the efficiency of our filter implementation within the TorchFX library, we conducted a benchmarking study comparing various Infinite Impulse Response (IIR) and Finite Impulse Response (FIR) filters against those provided by the SciPy library in conjunction with NumPy. Notably, libraries such as torchaudio and Julius were excluded from this analysis. The rationale for this exclusion lies in the fact that TorchFX is fundamentally based on torchaudio, making any direct comparison redundant, aside from minor overheads associated with wrapper classes. Additionally, other libraries like Julius, nnAudio, and Librosa were not included due to their lack of specificity in filter implementation. Julius, for example, only implements FIR filters, while both Librosa and nnAudio provide APIs that are excessively low-level and reliant on SciPy. Consequently, the only meaningful comparison was between SciPy and TorchFX.

\begin{table}[htbp]
    \centering
    \caption{Execution times for FIR filters (in seconds).}
    \label{tab:fir-times}
    \begin{tabular}{cccccc}
        \toprule
        Time (s) & Channels & GPU & CPU & SciPy \\
        \midrule
        5   & 1  & 0.014247 & 0.033221 & 0.020940 \\
        5   & 2  & 0.009029 & 0.032975 & 0.041547 \\
        5   & 4  & 0.001824 & 0.035984 & 0.087344 \\
        5   & 8  & 0.003292 & 0.034048 & 0.171197 \\
        5   & 12 & 0.004809 & 0.073933 & 0.251341 \\
        60  & 1  & 0.004731 & 0.464397 & 0.261466 \\
        60  & 2  & 0.009341 & 0.438233 & 0.525599 \\
        60  & 4  & 0.018447 & 0.469470 & 1.006187 \\
        60  & 8  & 0.053108 & 0.526290 & 2.054250 \\
        60  & 12 & 0.079765 & 1.030621 & 3.033699 \\
        180 & 1  & 0.013415 & 1.401418 & 0.777573 \\
        180 & 2  & 0.039879 & 1.340833 & 1.585332 \\
        180 & 4  & 0.079396 & 1.430514 & 3.129619 \\
        180 & 8  & 0.155036 & 1.579974 & 6.161899 \\
        180 & 12 & 0.231503 & 3.071292 & 9.509333 \\
        300 & 1  & 0.032687 & 2.341085 & 1.313260 \\
        300 & 2  & 0.064894 & 2.216330 & 2.598697 \\
        300 & 4  & 0.128962 & 2.362212 & 5.316212 \\
        300 & 8  & 0.257784 & 2.613295 & 10.421657 \\
        300 & 12 & 0.391072 & 5.047783 & 15.418619 \\
        600 & 1  & 0.064427 & 4.676215 & 2.635783 \\
        600 & 2  & 0.129250 & 4.567576 & 5.213626 \\
        600 & 4  & 0.254230 & 4.843752 & 10.236919 \\
        600 & 8  & 0.508825 & 5.179805 & 20.946205 \\
        600 & 12 & 0.785939 & 10.173771 & 31.074807 \\
        \bottomrule
    \end{tabular}
\end{table}

\begin{table}[htbp]
    \centering
    \caption{Execution times for IIR filters (in seconds).}
    \label{tab:iir-times}
    \begin{tabular}{cccccc}
        \toprule
        Time (s) & Channels & GPU & CPU & SciPy \\
        \midrule
        5   & 1  & 0.001074 & 0.006376 & 0.003344 \\
        5   & 2  & 0.001010 & 0.009064 & 0.006739 \\
        5   & 4  & 0.001054 & 0.015751 & 0.013377 \\
        5   & 8  & 0.001197 & 0.037838 & 0.027034 \\
        5   & 12 & 0.001199 & 0.066645 & 0.041763 \\
        60  & 1  & 0.282906 & 0.142985 & 0.043823 \\
        60  & 2  & 0.002125 & 0.191987 & 0.088169 \\
        60  & 4  & 0.001242 & 0.273665 & 0.342257 \\
        60  & 8  & 0.001255 & 0.443807 & 0.688531 \\
        60  & 12 & 0.001269 & 0.741386 & 1.038889 \\
        180 & 1  & 0.001324 & 0.435303 & 0.250236 \\
        180 & 2  & 0.001205 & 0.570106 & 0.502351 \\
        180 & 4  & 0.001259 & 0.805437 & 1.007327 \\
        180 & 8  & 0.001257 & 1.290896 & 2.017688 \\
        180 & 12 & 0.001281 & 2.139772 & 3.026279 \\
        300 & 1  & 0.001058 & 0.737175 & 0.418597 \\
        300 & 2  & 0.001053 & 0.948024 & 0.839463 \\
        300 & 4  & 0.001087 & 1.322156 & 1.683135 \\
        300 & 8  & 0.001244 & 2.115006 & 3.374017 \\
        300 & 12 & 0.001364 & 3.547629 & 5.054017 \\
        600 & 1  & 0.001063 & 1.465892 & 0.840227 \\
        600 & 2  & 0.001111 & 1.888490 & 1.686797 \\
        600 & 4  & 0.001095 & 2.641203 & 3.373638 \\
        600 & 8  & 0.001245 & 4.251984 & 6.886340 \\
        600 & 12 & 2.466622 & 7.079701 & 10.335053 \\
        \bottomrule
    \end{tabular}
\end{table}

Given TorchFX's dual capability to operate on both GPU and CPU platforms, our evaluation encompassed a tripartite comparison: TorchFX on CPU, TorchFX on GPU, and SciPy, focusing on the efficiency of FIR and IIR filter applications. Additionally, we conducted a benchmark to assess execution times across various interfaces as delineated in Section \ref{sec:design}. These interfaces include the pipe operator, a class extending nn.Module, and a sequence of filters concatenated in nn.Sequential. The experimental tests were conducted on an Alienware Aurora R11 1.0.8 system, running Linux Ubuntu 22.04 with kernel version 6.8.0-57-generic. This system was equipped with an NVIDIA GeForce RTX 3070 graphics card featuring 8GB of VRAM, CUDA version 12.4, and an Intel i9-10900KF CPU operating at 5.3 GHz, complemented by 32 GB of RAM.

The benchmarking process was structured, with evaluations based on the average execution time derived from 50 repetitions of identical tasks. For each algorithm, three distinct implementations\footnote{The benchmark code is publicly accessible at the GitHub repository.} were tested: one utilizing solely SciPy and NumPy on the CPU, another employing TorchFX exclusively on the CPU, and a third leveraging TorchFX with execution, where feasible, on the GPU. To optimize pipeline efficiency, our measurements focused solely on the execution time of filter application on the signal, deliberately excluding the computation of filter coefficients and the reading of the signal into memory from the timing assessments. The performance evaluation of the IIR and FIR filters was conducted by varying the input signal's duration, ranging from 5 seconds to 10 minutes, and altering the number of channels from 1 to 12. The processed signal comprised white noise, generated via the numpy.random.randn function, sampled at a frequency of 44100 Hz—a sampling rate commonly supported by contemporary computer audio cards.

\begin{figure*}
    \centering
    \includegraphics[width=\linewidth]{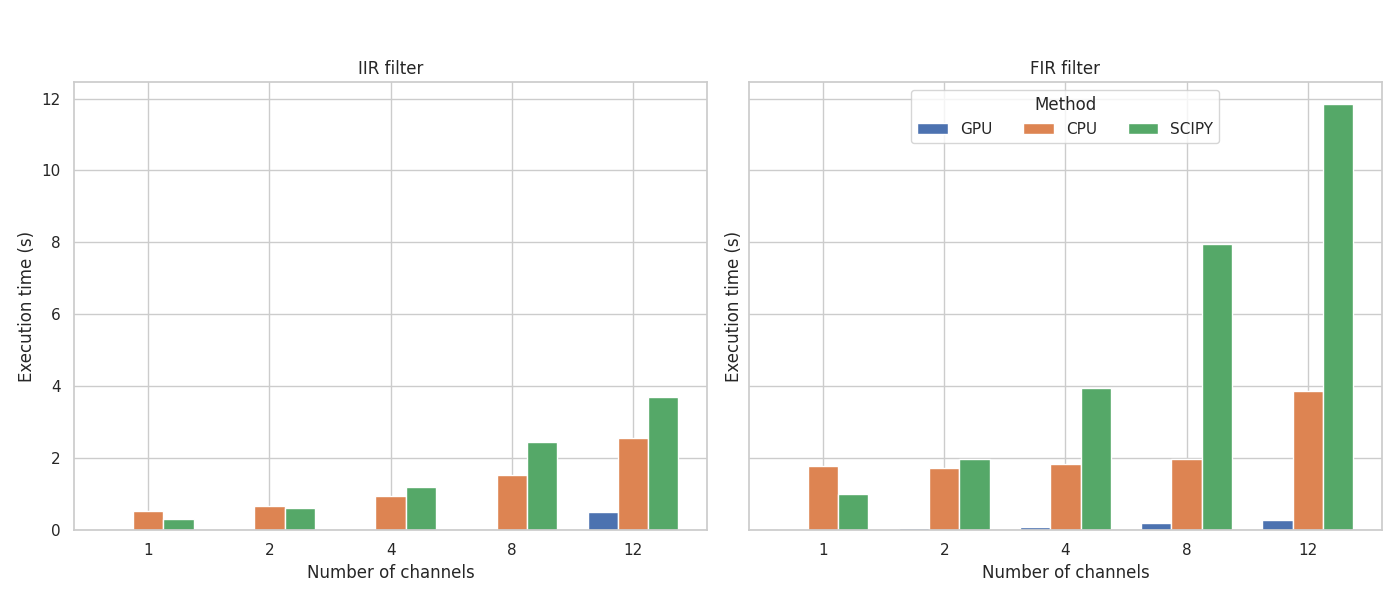}
    \caption{Comparison of the average execution time as a function of the number of channels for IIR and FIR filters.}
    \label{fig:filters-exe}
\end{figure*}

Figure \ref{fig:filters-exe} illustrates the results, showing that, overall, for single-channel and extremely short signals, filter application is more efficiently executed using SciPy. This efficiency is due to the underutilization of the GPU's parallel computing capabilities and the additional overhead incurred by VRAM memory transfers. However, as the number of channels increases, the execution time for the SciPy implementation escalates linearly, whereas TorchFX, both with and without GPU acceleration, demonstrates significantly enhanced efficiency, the exact performance numbers are displayed in table \ref{tab:fir-times} for FIR results and \ref{tab:iir-times} for IIR results.

Unsurprisingly, as the dimensions of the input increase, the CPU time of SciPy algorithms escalates dramatically. This phenomenon can be attributed to the inherent limitations of CPU processing, which struggles to efficiently handle larger datasets due to its sequential execution model. In stark contrast, the GPU implementation of TorchFX, leveraging its parallel processing capabilities, consistently maintains execution times well below one second, even for extensive input signals. This remarkable efficiency underscores the advantages of utilizing GPU acceleration for digital signal processing tasks, particularly when dealing with high-dimensional data.
It is also important to note that the CPU implementation of TorchFX remains competitive, particularly for larger signals. This is largely due to its optimization for multi-core processing, allowing it to effectively utilize all available CPU cores. As a result, even in scenarios where GPU acceleration may not be feasible, TorchFX’s CPU performance provides a viable alternative, ensuring that users can achieve efficient processing without the need for specialized hardware.
Furthermore, the benchmarking results highlight a critical aspect of digital signal processing: the choice of implementation can significantly impact performance. While SciPy excels in scenarios involving single-channel and short-duration signals, its performance deteriorates as the complexity of the input increases. In contrast, TorchFX demonstrates a more scalable approach, maintaining efficiency across varying input sizes and channel counts. This scalability is particularly beneficial for applications requiring real-time processing or handling of multi-channel audio, where the ability to manage increased computational demands without sacrificing performance is paramount.

The comparative analysis of interface efficiency was conducted on a signal sampled at 44100 Hz, encompassing 8 channels, with a duration of 2 minutes. This signal was synthetically generated in a random manner, and the average execution time was calculated from 50 algorithm repetitions using the Python timeit module. The filter series comprised a chain of two Butterworth filters and two Chebyshev1 filters. In this scenario, the SciPy implementation was found to be slower than all other implementations, which exhibited closely aligned performance metrics. Table \ref{tab:interface} demonstrates that the various implementations based on TorchFX achieved an average execution time of $1.56\pm0.01$ seconds, approximately half a second faster than the SciPy implementation.

\begin{table}[ht]
  \caption{\itshape Execution time as a function of the interface.}
  \centering
  \begin{tabular}{|c|c|}
    \hline
    Implementation & Time (s) \\\hline
    scipy & $2.0173$ \\
    nn.Module subclass & $1.5743$ \\
    nn.Sequential & $1.5487$ \\
    pipe operator & $1.5626$ \\\hline
  \end{tabular}
  \label{tab:interface}
\end{table}

\section{Conclusions}
\label{sec:conclusion}

In this article, we introduced TorchFX, a novel Python library designed to provide high-level programming interfaces to design complex filters in Python audio signal processing by leveraging the computational efficiencies of GPU acceleration. Built on the robust PyTorch framework, TorchFX provides an object-oriented interface that  simplifies the manipulation of audio signals. A key innovation of this library is the introduction of the pipe operator, achieved through the operator overloading of the OR operator, which allows users to intuitively chain multiple filters. This feature facilitates the seamless creation of complex filter chains, making the process both intuitive and straightforward. The library offers a comprehensive suite of FIR and IIR filters, with particular attention to the processing needs of multichannel audio files.

Despite the advancements and capabilities presented by TorchFX, it is important to acknowledge certain limitations that currently exist. One significant limitation is the compatibility with the available GPUs on the market. At present, TorchFX primarily supports CUDA, the programming language developed by NVIDIA for its hardware, which means that the library is compatible with NVIDIA GPUs. However, it does not currently support AMD or Intel GPUs. This limitation is a consequence of TorchFX being built on PyTorch, which is actively working to extend its support to these platforms. We are optimistic that TorchFX will become compatible with AMD and Intel GPUs in the future. Additionally, while there are ongoing projects exploring the use of Vulkan for GPU acceleration, these efforts are still in development and have not yet reached a stable state. Nevertheless, TorchFX remains compatible with CPUs, allowing it to be used on machines that do not have GPU capabilities.

Looking ahead, TorchFX is still under active development, and our roadmap includes several exciting enhancements. We plan to expand the library's interface by introducing additional filters and incorporating common DSP functionalities such as Fast Fourier Transform (FFT) and Short-Time Fourier Transform (STFT). These additions will further enhance the library's versatility and utility in various audio processing applications. Moreover, we are committed to enabling compatibility with real-time audio streams, which will open up new possibilities for live audio processing applications. This capability will be particularly valuable in fields such as music production, broadcasting, and interactive audio experiences.

In conclusion, TorchFX is a new tool for audio signal processing that integrates DSP techniques within an AI-development framework, aiming to simplify the implementation of advanced audio processing methods. As the library continues to evolve, it aspires to become a valuable resource for researchers and developers in the audio processing domain, contributing to advancements in GPU-accelerated audio processing.

\bibliographystyle{IEEEbib}
\bibliography{references}

\end{document}